\documentclass[12pt]{article}

\usepackage{scicite}

\usepackage{times}

\usepackage{amsmath}
\usepackage{amsfonts}
\usepackage{amssymb}
\usepackage{graphicx}
\usepackage{booktabs}
\usepackage{multirow}

\usepackage{xcolor}
\usepackage[hidelinks]{hyperref}

\definecolor{darkorange}{rgb}{1, 0.55, 0}

\newenvironment{sciabstract}{%
\begin{quote} \bf}
{\end{quote}}

\title{Which products activate a product? An explainable machine learning approach}

\author
{Massimiliano Fessina,${}^{1}$ Giambattista Albora,${}^{2,3\ast}$ Andrea Tacchella,${}^{4}$ Andrea Zaccaria${}^{5,2}$\\
\\
\normalsize{${}^{1}$IMT School for Advanced Studies, Lucca, Italy}\\
\normalsize{${}^{2}$Centro Ricerche Enrico Fermi, Rome, Italy}\\
\normalsize{${}^{3}$Sapienza University of Rome, Italy}\\
\normalsize{${}^{4}$Joint Research Centre, Seville, Spain}\\
\normalsize{${}^{5}$Institute for Complex Systems, CNR, UOS Sapienza, Rome, Italy}
\\
\normalsize{$^\ast$To whom correspondence should be addressed; E-mail:  alboragiambattista@gmail.com}
}

\date{}

\begin{document} 


\baselineskip24pt


\maketitle

\begin{sciabstract}
Tree-based machine learning algorithms provide the most precise assessment of the feasibility for a country to export a target product given its export basket. However, the high number of parameters involved prevents a straightforward interpretation of the results and, in turn, the explainability of policy indications. In this paper, we propose a procedure to statistically validate the importance of the products used in the feasibility assessment. In this way, we are able to identify which products, called \textit{explainers}, significantly increase the probability to export a target product in the near future. The explainers naturally identify a low dimensional representation, the Feature Importance Product Space, that enhances the interpretability of the recommendations and provides out-of-sample forecasts of the export baskets of countries. Interestingly, we detect a positive correlation between the complexity of a product and the complexity of its \textit{explainers}.
\end{sciabstract}

\section*{Introduction}
The mechanisms underlying economic development \cite{acemoglu2012introduction} are among the most studied branches of economics since the work of Adam Smith \cite{smith1887inquiry}. However, the identification of its determinants remains an open problem, despite the flourishing of different models and interpretations \cite{helpman2009mystery}; in particular, standard theories, based in aggregated measures of production inputs, have limited capacity to predict growth and to recommend specific industrial policies \cite{barro1989economic}. The line of research based on the works of \cite{penrose2009theory,teece1994understanding,sutton2012competing} moves from the presence of the so called capabilities, the set of endowments countries have and that permit their industrialization and developments. Capabilities are, in practice, hard both to define and measure, since in principle they could span from human capital, to infrastructures, government and so on.
The solution proposed in \cite{hausmann2007you} is to infer them from the export baskets, i.e. the diversification structure provided by the set of products exported by the country under investigation. This idea opened up the possibility to apply techniques and methodologies borrowed from physics and network science, which go under the name of economic complexity \cite{hidalgo2009building,hidalgo2007product,tacchella,sbardella2018role}. In particular, the approach discussed by Tacchella et al.\cite{tacchella} aims at building a synthetic measure of the Fitness of a country, which is able to forecast the GDP growth with a precision higher than the state of the art methodologies\cite{tacchella2018dynamical}. However, this approach provides a \textit{global} picture of the country, while a more detailed analysis is often needed in order to provide specific industrial recommendations 
\cite{lin2020african,pugliese2020economic}. In this perspective, a number of papers built networks whose nodes are products and links are given by their similarity, proxied by their co-occurrences in the export baskets of countries \cite{hidalgo2007product,zaccaria2014taxonomy,pugliese2019unfolding}. In such a way, two products can be defined as close in the sense that they share many of the capabilities needed in order to export them in a competitive way. Co-occurrences based approaches have however a low predictive performance, and this fact favors machine learning approaches as better tools to measure relatedness both at country \cite{tacchella2021relatedness,albora2021product,che2020intelligent} and firm level \cite{albora2022machine,Straccamore_2022}. In \cite{zaccaria2014taxonomy,tacchella2016build,saracco2015innovation}, the authors proposed approaches to explicitly model the relationship among products, capabilities, and development. These frameworks naturally lead to the concepts of product progression \cite{zaccaria2014taxonomy,zaccaria2018integrating,albora2021product} and arrow of development \cite{o2021productive}: the relationship between products is often not undirected, or symmetric, as in the product space \cite{hidalgo2007product}, but \textit{directed}: countries starts their development from simple products and gradually enter in more sophisticated markets, following well defined paths of development \cite{zaccaria2014taxonomy}. Obviously, the identification of the specific products enabling countries to competitively export a given target product is a key element to design industrial policies and strategic patterns of development. Despite the importance of this investigation, a specific analysis was missing because of the lack of suitable tools and algorithms able to successfully forecast the export of countries. However, thanks to the introduction of machine learning in the economic complexity analysis \cite{tacchella2021relatedness}, the tools at disposal reached a maturity such that this investigation can start providing concrete and scientifically validated results. This is the aim of the present paper: to provide an algorithmic approach based on a highly predictive machine learning method to measure the importance of single products and sectors for a country to export a specific target product in a given amount of years. The link between starting and target products will be quantified by using the feature importance, a key tool of supervised machine learning algorithms which allows a clear interpretation of the outputs.\\

\section*{Results}
\subsection*{The predictive framework}
Our aim is to understand which products enable a country to export a given target product. To do so, we investigate the mechanisms underlying a machine learning based prediction approach \cite{tacchella2021relatedness}. Such approach considers the competitiveness level of each country's export on each product as \textit{features} \cite{albora2021product}: obviously, some products will be dominant in the forecast exercise, while others will be practically irrelevant. The \textit{feature importance} \cite{geron2019hands,shalev2014understanding,albora2021product} will be our statistically validated measure of the ability of a product to activate another product. It is obviously of key importance to adopt a framework which has an excellent forecasting power. The approach discussed here, based on the Random Forest (RF) algorithm \cite{breiman2001random}, outperforms the networks of co-occurrences \cite{tacchella2021relatedness} as well as other supervised machine learning algorithms \cite{albora2021product}, also when other data typologies are considered \cite{tacchella2021relatedness,albora2022machine,Straccamore_2022}. Here we briefly summarize the predictive framework. Full details are provided in the Methods section.\\
The predictive task is represented by the out-of-sample forecast of the appearance of new links in the country-product temporal network\cite{albora2021product}. At a given year $y$, the network represents whether country $c$ exports product $p$ in a competitive way or not. Mathematically, it is identified by the adjacency matrix $M$ whose elements are

\begin{equation}
  M_{cp}(y) =
    \begin{cases}
      1 & \text{if}\,\,RCA_{cp}(y) \geq 1\\
      0 & \text{otherwise}
    \end{cases}    
    \label{eq:mcp}
\end{equation}

\noindent where $RCA_{cp}(y)$ is the \emph{Revealed Comparative Advantage} \cite{balassa1965trade}, defined in the Methods section. Given the knowledge of the network in a certain time interval, the RF algorithm can be trained in an appropriate cross-validated framework to make out-of-sample predictions on the matrix $M_{cp}(y+\delta)$, starting from the knowledge of $M_{cp}(y)$, that is, which country exports which product. In particular, a different RF model is trained for each product, and the other products are used as inputs, or features; in such a way, the RF learns from the past which products are usually associated with the target product.\\
In the present study we cover the time span 1996-2018, and choose a time interval $\delta=5$ years: the algorithm is trained on years 1996-2013 to make predictions on 2018. The number of countries is 169, while products are classified according to the Harmonized System (HS) 1992, which has a hierarchical structure: products can be aggregated in 97 different sectors (2-digit code level), or split into 5040 detailed products (6-digit code level). The size of the matrix $M$ will change accordingly. In the Materials and Methods section we provide more details about the data and the construction of the predictive model. 

\subsection*{Feature importances}

The interpretation of the predictions provided by the RF starts with the quantification of the importance the algorithm assigns to each feature (i.e., product) during the training procedure. In our setting, the goal is to forecast whether a product $p$ at the 6-digit level will be exported by a country in year $y+\delta$, knowing in which of the 97 2-digit sectors the country is active in year $y$ (always in the RCA sense). The 2-digit sectors are hence used as features and their importance is a measure of how much the activity of a country in each sector (i.e. export or non-export) is informative in order to determine if it will export the 6-digit product $p$ after $\delta$ years.

\begin{figure}[ht]
\centering 
\includegraphics[width=0.8\textwidth,height=0.32\columnwidth]{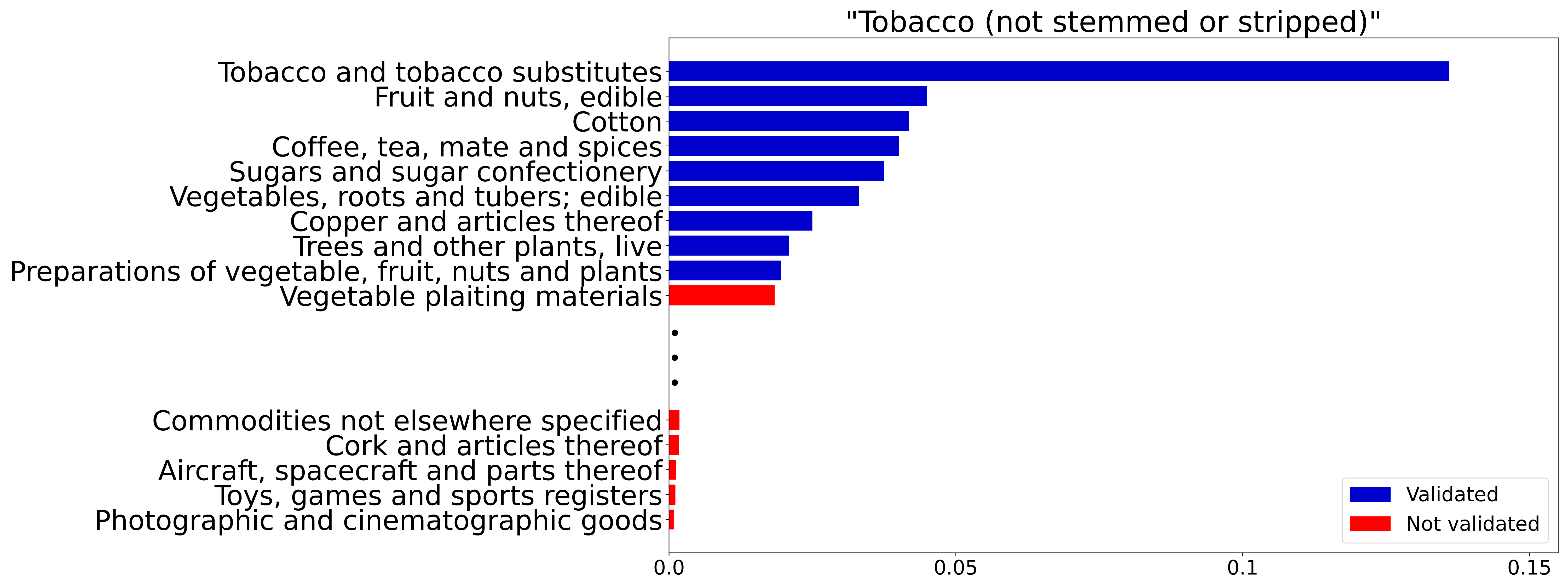}
\includegraphics[width=0.8\textwidth,height=0.32\columnwidth]{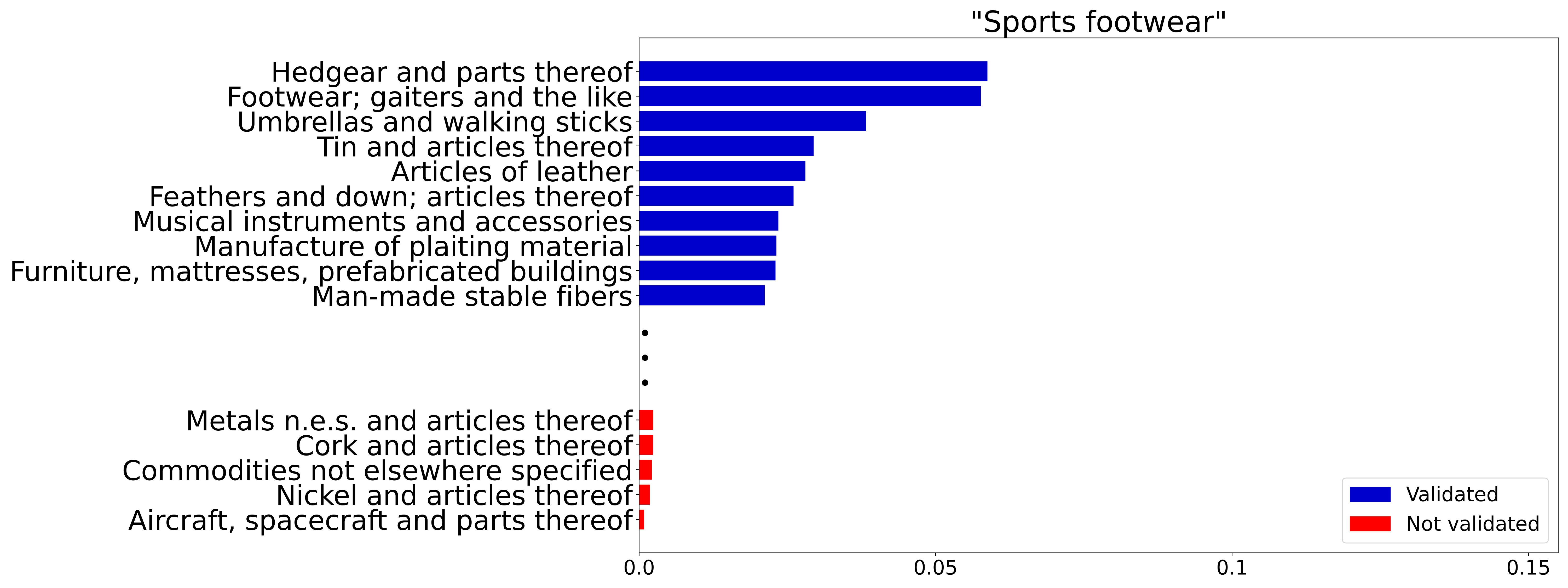} 
\includegraphics[width=0.8\textwidth,height=0.32\columnwidth]{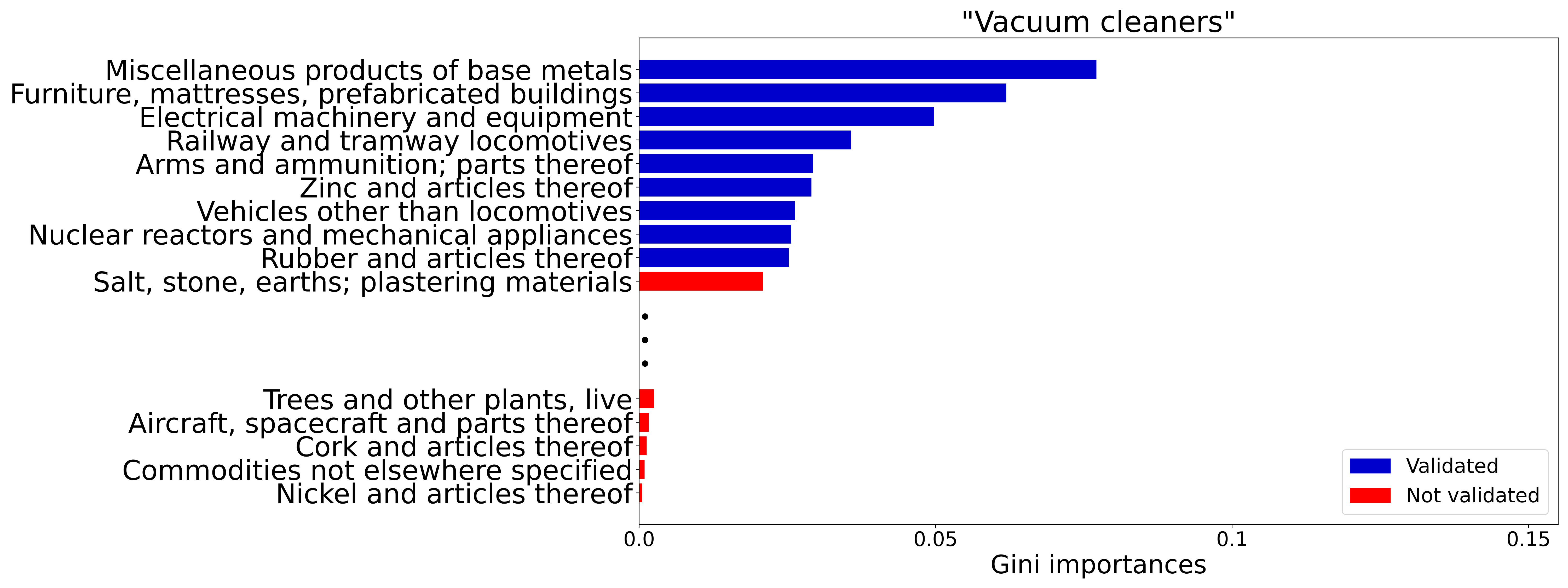} 
\caption{\textbf{Feature importances barplots.} Three examples of feature importances barplots; from top to bottom, "Tobacco", "Sports footwear", and "Vacuum cleaners". We report the top 10 and bottom 5 features: for all three products, the method validates a set of very reasonable features. Noticeably, in all three cases, the aggregated sectors to which the products belong are present among the explainers: "Tobacco and tobacco substitutes" (code \textbf{24}) for "Tobacco (not stemmed or stripped)" (code \textbf{240110}), "Footwear; gaiters and the like" (code \textbf{64}) for "Sports footwear" (code \textbf{640411}) and "Electrical machinery and equipment (code \textbf{85}) for "Vacuum cleaners" (code \textbf{850910}).}
\label{fig:gini_barplot}
\end{figure}

The quantification of feature importances is obtained using the \emph{Gini importance} \cite{breiman1984cart} (or \emph{mean impurity decrease}), a Random Forest-specific measure assigning to the features importance values summing up to 1.
Starting from the raw values, we performed a suitable statistical validation procedure, computing the corresponding \emph{p-values} and imposing a validation threshold of 95\%: such validation is based on the computation of the \emph{null importances}, i.e. the importance values the algorithm assigns to each variable after its association with the target vector is broken (see Materials and Methods for a detailed description of the procedure)\cite{altmann2010permimp}. Only the statistically validated importances are kept, while the others are put equal to zero. Hence we obtain, for each of the 5040 predicted products, a vector containing the validated importance measures for the 97 aggregate productive sectors. We call the products retaining a significant importance value \textit{explainers}: these products enhance the probability of a country to competitively export the target product as they signal the presence of the capabilities needed for it. In Figure \ref{fig:gini_barplot} we report the barplots of the feature importances for the products "Tobacco (not stemmed or stripped)" (code \textbf{240110}), "Sports footwear" (code \textbf{640411}) and "Vacuum cleaners" (code \textbf{850910}), showing the 10 most important and the 5 least important sectors. The colors represents whether the feature importance has been statistically validated (blue), or not (red). In all three cases we can notice how the explainers can be intuitively related to the products: e.g. the 2-digit sectors to which the 6-digit products belong are correctly recovered among the explainers (respectively, "Tobacco and tobacco substitutes", code \textbf{24}, "Footwear; gaiters and the like", code \textbf{64}, and "Electrical machinery and equipment", code \textbf{85}). This represent a first qualitative test of the ability of the implemented methodology to recover significant correlations between productive sectors and products, as learned by Random Forest in its training procedure.

\clearpage

\subsection*{Feature Importance Product Space} 

\begin{figure}[ht]
\centering 
\includegraphics[width=\textwidth]{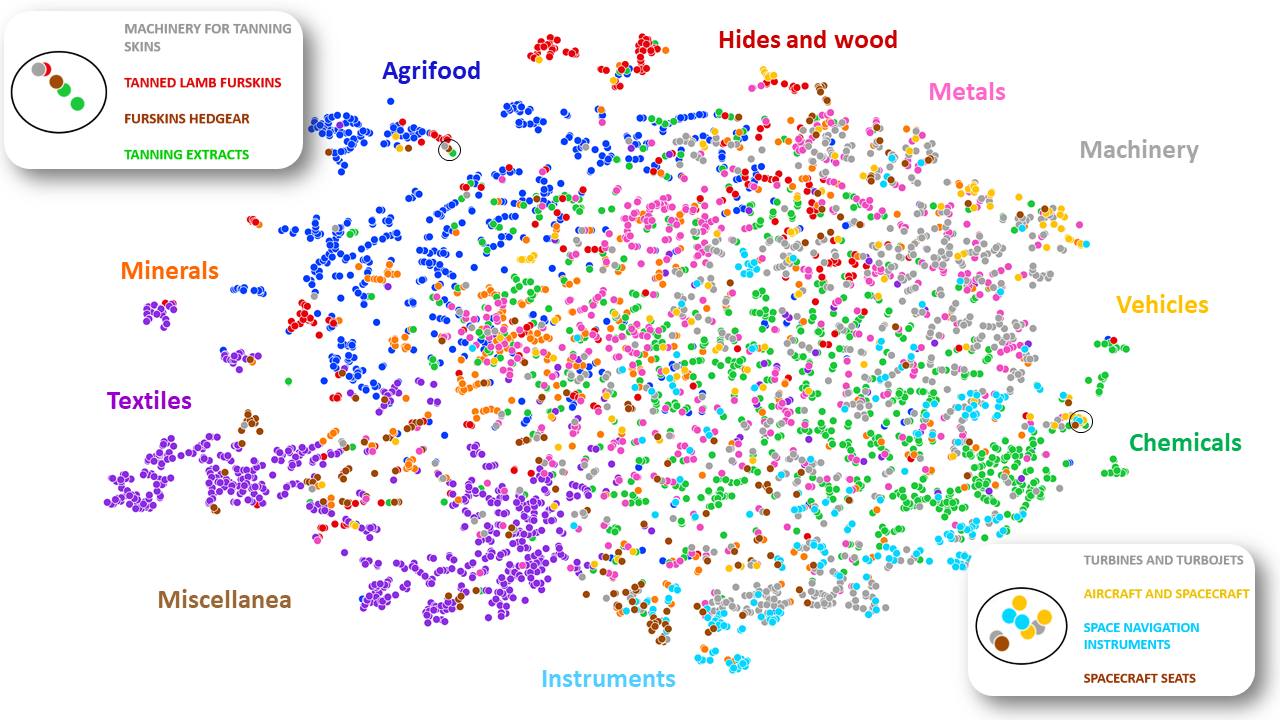}
\caption{\textbf{Feature Importance Product Space (FIPS).} The FIPS is a low-dimensional representation of the feature importance vectors: each dot is a product, identified by its explainers. The colors correspond to ten productive macro-categories. The structure of the space is heterogeneous, with some relatively isolated categories of products (e.g. Textiles and Agrifood), and areas occupied by a mixing of different kinds of products (e.g. the right side of the figure, where there is a mixing of Machinery, Vehicles, Chemicals and Instruments). The two insets are zooms which testify the ability of the FIPS to group together similar products: the first (upper-left side) is composed of four different kind of products, all related to furskins; the second (lower-right side) groups together products belonging to the spacecraft industry.}
\label{fig:fips}
\end{figure}

The Gini importance vectors can be interpreted as high-dimensional representations for the products, like word embeddings \cite{mikolov2013efficient} in natural language processing \cite{jurafsky2000speech}.
Indeed, they contain information about the productive background that the Random Forest algorithm recognizes as necessary or highly predictive for their future export. Hence, the distance between such vectors can be used as a proxy for products' similarity: two products whose Gini importance vectors are close need a similar presence/absence pattern of capabilities in order to be competitively exported. \\
To test this hypothesis we projected the 97-dimensional vectors on a 2-dimensional continuous space, using the dimensionality reduction algorithm t-SNE \cite{van2008visualizing}: the result, which we call Feature Importance Product Space (FIPS), is reported in Figure \ref{fig:fips}. Here, each dot represents a 6-digit product, and the colors correspond to ten aggregate macro-categories (see Supplementary Information section S3). The structure of the FIPS is heterogeneous, with clusters of products belonging to single categories, as for Agrifood and Textiles (left side of the plot) and regions with the superposition of different product categories, as in the right side of the plot, where there is a mixing of Machinery, Vehicles, Chemicals and Instruments. This differentiation can be traced back to the complexity of products making up different sectors: less sophisticated sectors tend to be more distinguishable, as they need few capabilities, and therefore share similarity patterns with a smaller set of other products (see Materials and Methods and Supplementary Information section S1 for further analyses). On the contrary, high-complexity products share large portions of the respective production lines and supply chains \cite{taglioni2016making}. To highlight the ability of the space to identify the similarity of products even if they originally belong to different productive categories, we pinpointed two small clusters: the first (upper-left side of the figure) groups products related to the fur manufacture; the second (lower-right side of the figure) puts together different typologies of products, all related to the spacecraft industry.

\clearpage

\subsection*{Predicting products' appearances with FIPS}

The ability to make out-of-sample forecasting on the country-product network represents the natural field to test and compare the validity of relatedness measures \cite{albora2021product}.
Therefore, in order to quantify the goodness of the FIPS reconstruction and the amount of information it brings, we use it to predict the appearance of new products in $M_{cp}(2018)$, employing a density-based approach \cite{hidalgo2007product}. In other words we predict that countries will become competitive in new products which are close in the FIPS space to other products in which the country is already competitive. Practically predictions on a single product, for every country, are based on the amount of already exported products, each weighted by its link with the target product. In Table \ref{tab:fips_perf} we report the predictive performance of the FIPS, together with the performance of the Random Forest from which the FIPS was built, and with the temporal auto-correlation baseline represented by $RCA_{cp}(2013)$. The adopted performance metrics are (see Materials and Methods for a detailed discussion):
\begin{itemize}
\item Best F1 Score: the F1-score \cite{van1974foundation}, i.e. the harmonic mean of Precision and Recall \cite{powers2011evaluation}, computed for the decision threshold that maximizes its value;
\item AUC-ROC \cite{hanley1982meaning}: the area under the Receiving Operator Characteristic curve;
\item mP@10: the average, over the countries, of the Precision score on the top 10 predicted products.
\end{itemize}

The performances have not been computed on the full matrix $M_{cp}(2018)$, but on the so called \emph{activations}, i.e. those elements showing a value $RCA_{cp}(y)<0.25$ for $y\in[1996-2013]$: as noted in previous works \cite{tacchella2021relatedness,albora2021product}, this guarantees that the measured performance is indicative of the actual ability of the models to forecast genuine economic development (i.e. the appearance of a new product in the export basket of a country), rather than relying on the strong temporal auto-correlation of the country-product network.

The scores show that the FIPS performs better than the original RF for both Best F1 Score and mP@10, while achieving a lower value of AUC-ROC: this result is extremely relevant, as it implies that the FIPS has an overall higher forecasting power than the Random Forest from which it was built, while providing a clear interpretability of its predictions in terms of similarity relationships between products. Moreover, the AUC-ROC metrics is the least reliable, due to the strong class imbalance of the dataset (see \cite{albora2021product} and Materials and Methods). As a reference we also report the comparison with the RCA baseline, which outperforms both regarding the Best F1 Score. This is due to the different granularity of the inputs: the 6-digit products for RCA, the 2-digit aggregated sectors for FIPS and RF. Note that when the RF is trained at 6-digit it easily overcomes the RCA baseline \cite{tacchella2021relatedness,albora2021product}, but the performance of RCA still remains an important benchmark as it has been shown to perform substantially better than co-occurrence based approaches \cite{tacchella2021relatedness}. 

\begin{table}[!hb] 
\centering 
\medskip 
\begin{tabular}{cccc}
\toprule
& \textbf{Best F1 Score} & \textbf{AUC-ROC} & \textbf{mP@10}\\
\midrule
\textbf{Random Forest} & 0.033 & \textbf{0.698} & 0.041\\
\textbf{FIPS} & 0.035 & 0.671 & \textbf{0.047}\\
\textbf{RCA(`13)} & \textbf{0.037} & 0.592 & 0.039\\
\bottomrule
\end{tabular}
\caption{\textbf{Comparison of prediction performances of FIPS, Random Forest and RCA baseline.} The values of the performance metrics show that the FIPS performs overall better than the Random Forest, showing a lower value only of AUC-ROC: this result is very important, as it guarantees that the FIPS not only provides a fully interpretable predictive model in terms of products' similarity relationships, but retains and even overcomes the predictive power of the Random Forest it was built from. The RCA baseline has the highest Best F1 Score, being trained on the full 6-digit level data.}
\label{tab:fips_perf}
\end{table}

We expect, however, that the FIPS is uncovering fundamental capability-based explanations, that are sensibly different from the autocorrelation signal expressed by the RCA, and this can't immediately be seen from the forecasting performance scores.
In order to assess the additional information carried by the FIPS with respect to the temporal auto-correlation of the network, we decided to plug both the prediction score on $M_{cp}(2018)$ by FIPS and the $RCA_{cp}(2013)$ as variables into a logistic regression whose dependent variable is the possible activation of a product. The logit model is trained on the \emph{activations} ($RCA_{cp}(y)<0.25$ for $y\in[1996-2013]$) in an appropriate cross-validated setting, to make out-of-sample predictions on $M_{cp}(2018)$ (see Materials and Methods). The results, reported in Table \ref{tab:logit_model}, confirm the validity of the information carried by the FIPS as complementary with respect to the network auto-correlation in two ways. First of all, the logit model trained on both FIPS and RCA has the highest value of Pseudo $R^{2}$; secondly, this model displays a better predictive performance with respect to both logit models trained on $RCA_{cp}(2013)$ and FIPS alone. 
We further compare it with the prediction accuracy provided by $RCA_{cp}(2013)$, RF, and FIPS alone (i.e. without being used as variables in a logistic regression), showing that the \textbf{FIPS + RCA} logit model has the highest Best F1 Score, and it trails only Random Forest for the AUC-ROC score.

\begin{table}[!hb] 
\centering 
\medskip 
\begin{tabular}{cccc|ccc}
\toprule
& \multicolumn{3}{c}{\textbf{Logit\,\,model}} & \multicolumn{3}{c}{\textbf{Direct\,\,predictions}} \\
\hline
\midrule
& \textbf{RCA} & \textbf{FIPS} & \textbf{FIPS + RCA} & \textbf{RCA} & \textbf{RF} & \textbf{FIPS} \\
\midrule
\textbf{RCA}  & 9.015*  $\pm$ 0.374 &  & 7.012* $\pm$ 0.406 & & & \\
\textbf{FIPS} &  & 4.782* $\pm$ 0.166 & 4.018* $\pm$ 0.177 & & & \\
\textbf{Constant} & -5.223* $\pm$ 0.023 & -5.315* $\pm$ 0.024 & -5.391* $\pm$ 0.025 & & & \\
\textbf{Pseudo\,\,$R^{2}$} & 0.014 & 0.019 & \textbf{0.027} & & & \\
\textbf{AUC} & 0.584 & 0.659 & 0.685 & 0.592 & \textbf{0.698} & 0.669 \\
\textbf{BestF1} & 0.035 & 0.034 & \textbf{0.039} & 0.037 & 0.033 & 0.035\\
\bottomrule
\end{tabular}
\caption{\textbf{Logistic regression carried out with FIPS and $\mathbf{RCA_{cp}(2013)}$ to predict $\mathbf{M_{cp}(2018)}$.} The values of Pseudo $R^{2}$ show that the information carried by the FIPS and the RCA baseline are complementary, as the logistic regression trained on both models shows the highest value. This is confirmed by the performance metrics, as the latter shows a performance higher than both FIPS and RCA, when used individually (both directly and in a logistic regression setting) to make predictions; Random Forest anyway retains the highest AUC-ROC value. All performances are computed on the new products activations defined by $RCA_{cp}(y)<0.25$ for $y\in[1996-2013]$. The asterisks indicate that all the coefficients are statistically validated within a $99.9\%$ significance threshold.}
\label{tab:logit_model}
\end{table}

\subsection*{Feature importance and products' complexity}

\begin{figure}[ht]
\centering 
\includegraphics[width=\textwidth]{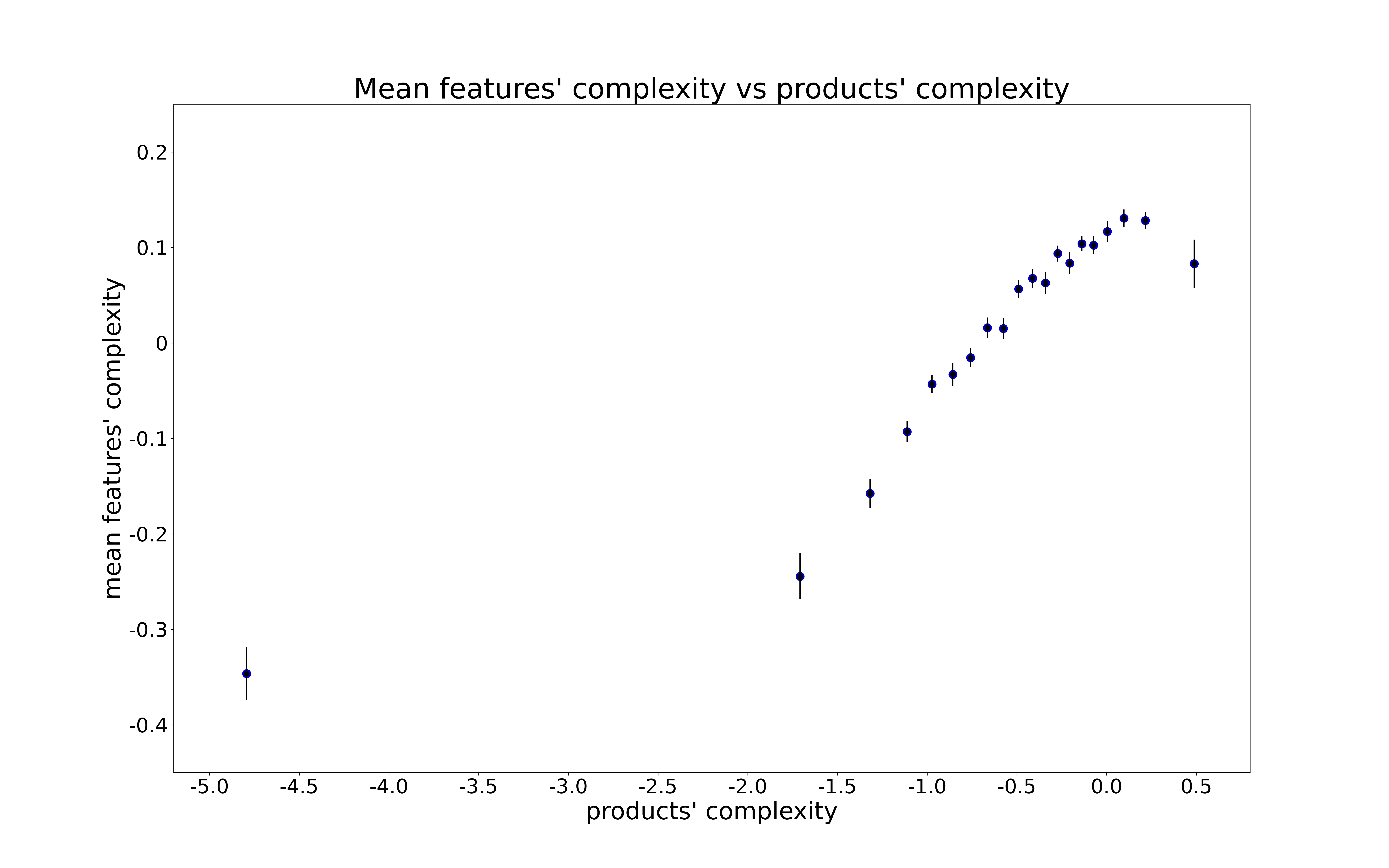}
\caption{\textbf{Mean complexity of explainers versus the mean complexity of target products.} Non parametric regression of the mean complexity of features with respect to the complexity of the target products. The values of complexity are computed as averages of the corresponding logarithms in the time span 1996-2013. The target products are grouped into 20 bins of 252 products each, for which we show the average mean complexity of the respective features and the corresponding standard error.}
\label{fig:fimp_compl_vs_compl}
\end{figure}

Another key assessment of this study is the unveiling of a connection between the feature importance vector of a product and its complexity. The complexity of a product, defined by the \emph{Economic Fitness and Complexity} algorithm \cite{tacchella,tacchella2013economic}, is a non-monetary indicator related to the level of industrial sophistication needed to competitively export it on the global market. As such, we expect it to be connected to the nature of the \textit{explainers} obtained for a product, as they represent the productive sectors recognized by our model as necessary for the future export of the product: the more complex a product, the more complex we expect the corresponding explainers to be. Since we train our Random Forest models using data in the time span 1996-2013, the complexities of products were computed as the average of the annual (log-) complexities in the same interval.
The visualization of the average complexity of the validated features versus the complexity of the corresponding target products (Fig. \ref{fig:fimp_compl_vs_compl}) confirms this idea:  more complex products need, on average, more complex features in order to be competitively exported. This finding confirms that the production lines of highly sophisticated products are deeply entangled among themselves \cite{taglioni2016making,angelini2018complexity}.

\section*{Discussion}

Relatedness \cite{hidalgo2018principle} is a central topic of the economic complexity approach and a key element for investment decisions and policy makers \cite{pugliese2020economic,lin2020african}. The idea is to empirically measure how close a country is to exporting a new product, that is to assess the feasibility of such a strategy. By comparing the predicting performances of different methodologies, recent studies \cite{tacchella2021relatedness,albora2021product} showed that machine learning algorithms such as Random Forest (RF) provide the state-of-the-art assessment of relatedness; here the features of this supervised machine learning approach are the products which are present or absent in the export basket of countries. The cost of a better prediction and relatedness assessment is, however, a reduced interpretability of the results, at least with respect to the traditional, network-based approaches \cite{hidalgo2007product,zaccaria2014taxonomy}. Nevertheless, having a visual representation of the diversification dynamics of countries, as well as
knowing which products are the most relevant to activate (or to \textit{explain}) the export of a new product is essential in order to inform industrial policies and to understand the different patterns of economic development.\\ 
In this study, we address the problem of the black box nature of the RF algorithm by proposing a methodology to extract information on the relevance of each input feature (a 2-digit sector) as a predictor of the future export of each of the 5000 possible target products at 6 digits. The starting point is the construction of a predictive model for the possible future export of each target product, based on the training of a RF algorithm. We then apply a procedure to statistically validate the Gini Importance of the single input features; in this way we are able to identify the \textit{explainers}, the key products needed by a country to competitively export a target product in the near future.
The importance the algorithm assigns to each input feature for every target product can be arranged in a  97-dimensional feature importance vector, which represents a highly dimensional embedding of the about 5000 target products. By means of the \emph{t-SNE} algorithm \cite{maaten2008visualizing}, we project such vectors on a 2-dimensional continuous space we call Feature Importance Product Space (FIPS). Here each point represents a product, and the closeness between points indicates that the corresponding products are similar in the sense that they share most of the explainers needed for their export. As such, this approach is closer to the theoretical approach discussed in the seminal papers by Teece et al. \cite{teece1994understanding,teece1997dynamic}, in which the capability overlap between products is detected a posteriori by counting their co-occurrences, an approach known in the complexity field as the Product Space \cite{hidalgo2007product}. Here, instead, the proximity is assessed by comparing which \textit{input} sectors are needed for the target products; similar explainers clearly imply similar capabilities.\\
By means of a density-based approach we employ the FIPS to forecast future exports, revealing that it provides better predictions than the RF it is built from. In order to test the additional information carried by the FIPS with respect to the strong temporal auto-correlation of the network, we used it as input variable into a logistic regression, together with the Revealed Comparative Advantage (RCA) of countries on products: besides confirming the significance of the FIPS as a predictor of future exports, this model displays an even better performance, overcoming the strong benchmark model given by RCA itself.\\
These results confirm the validity of our approach, as the FIPS represents a forecasting model on countries' future exports which retains (and even overcomes) the predictive power of the black-box algorithm it is based on, while allowing for a clear interpretability of its predictions, typical of the low performing network-based approaches. Aside from the forecasting power, the FIPS correctly captures information about the complexity of products, by inserting the most sophisticated sectors into dense clusters and isolating the less complex products. These findings pave the path to characterize the capabilities needed to be competitive in the export of complex products.\\
To conclude, we observe that using the 97 2-digit aggregated sectors as features reduces the predictive power of the RF with respect to using the 6-digit 5040 products. This choice for the setting of the study was due the computational time needed for the construction of the model, which would have been even higher with 5040 features. However, an optimization of the machinery in order to perform the analysis also for 4- and 6-digits features is currently under study, and will be the subject of future works. 

\section*{Materials and Methods}

\subsection*{Data}

The starting data used in this study is gathered by UN-COMTRADE and available upon subscription on the website https://comtrade.un.org. UN-COMTRADE provides the annual bilateral export flows between countries at the 6-digit product level.
Products are classified according to the Harmonized Commodity Description and Coding System, in its 1992 version (HS-1992): each product is identified by a 6 digits code, where each couple of digits refers to a different aggregation level. The total number of products ranges from 97 at the 2 digit level (aggregated sectors), to 5040 at the 6 digit level (detailed products).

Since importers' and exporters' declarations not always coincide, a Bayesian reconciliation procedure \cite{mazzilli2021reconstruct} is performed on data, leading to the definition of the annual export matrices $E_{cp}(y)$. Each element corresponds to the export volume realized by country $c$, for product $p$, in year $y$. The total number of countries is 169, and the covered time span is 1996-2018.

Following the standard procedure in the economic complexity literature \cite{hidalgo2007product,tacchella}, we compute the Revealed Comparative Advantage \cite{balassa1965trade}:

\begin{equation}
    RCA_{cp}(y) = \frac{E_{cp}/\sum_{p'}E_{cp'}}{\sum_{c'}E_{c'p}/\sum_{c'p'}E_{c'p'}}.
\end{equation}
This economic indicator measures the ratio between the weight that the export of a product $p$ has for country $c$ and the weight it has on the global market. In this way, we can filter out the size effects of both countries and industrial sectors. Finally, imposing a threshold equal to $RCA_{cp}=1$, distinguishing whether country $c$ is a competitive exporter of product $p$ in year $y$, we obtain the binary adjacency matrices $M_{cp}(y)$, as described in equation \ref{eq:mcp}.

\subsection*{Random Forest}

In order to forecast the export of countries, we train a supervised machine learning algorithm. In particular, we train one model for each target product; being the answer binary, we adopt a classification algorithm. Random Forest \cite{breiman2001random} is an ensemble method based on the aggregation of several decision trees \cite{james2013intro}: the final prediction of the algorithm is given by the average of the predictions made by the single trees. 

The Random Forest has been shown \cite{albora2021product} to be the top performing algorithm, together with XGBoost \cite{chen2016xgboost}, for our predictive task, which is discussed in detail in the next section. We point out that XGBoost is practically unfeasible for the specific investigation discussed here because of the needed computational effort. Moreover, the extraction of the feature importances is much more direct in the case of Random Forest.\\
In this study we made use of the \emph{Python} implementation provided by the library \emph{scikit-learn}\footnote{https://scikit-learn.org/stable/modules/generated/sklearn.ensemble.RandomForestClassifier}, which makes use of the \emph{CART} version of the algorithm \cite{breiman1984cart}. The hyperparameters\cite{geron2019hands} were set to their default values, a usual choice given the relative stability of the predictive performance
\cite{genuer2008random,fernandez2018learning,probst2019hyperparameters}.

\subsubsection*{Predictive Model}

The aim of the application of the Random Forest algorithm \cite{breiman2001random} to the country-product network is to build a predictive model able to forecast the export baskets of countries after $\delta$ years, given the knowledge of their present export baskets. This means predicting the structure of the network $M_{cp}(y+\delta)$ starting from $M_{cp}(y)$.

This is realized through the construction of a single model for each target product $p'$, performing a binary classification task. Given the knowledge of the network in the time span $[y_{0},y_{f}]$, such model is trained on the set:

\begin{itemize}
    \item $X_{train}$ = $\bigl\{M_{cp}(y), y\in[y_{0},y_{f}-2\delta]\bigr\}$
    \item $y_{train}$ = $\bigl\{M_{cp'}(y), y\in[y_{0}+\delta,y_{f}-\delta]\bigr\}$
\end{itemize}
and, in this process, learns which export baskets in $X_{train}$ are associated to the countries exporting or not exporting $p'$ ($y_{train}$).
The test set is defined in a similar way:

\begin{itemize}
    \item $X_{test}$ = $M_{cp}(y_{f}-\delta)$
    \item $y_{test}$ = $M_{cp'}(y_{f})$.
\end{itemize}
In this way we make sure the test is performed on completely unforeseen data, and prevent the algorithm from having any information about the structure of the network in years $y>y_{f}-\delta$ during the learning phase. The data relative to different years is stacked together vertically: in this perspective each country in each year represents an observation, the export baskets for all products its features, and the possible export of $p'$, $\delta$ years later, the corresponding class.
Putting together the predictions provided for all products, we recover the full matrix of predictions whose elements $S_{cp}(y_{f})$, can be tested against $M_{cp}(y_{f})$. It is to be noted that the prediction on a single element $S_{cp}(y_{f})$ is a probability value between 0 and 1, to be binarized with the choice of a threshold in order to be compared to the empirical element $S_{cp}(y_{f})$.

So, for each product $p'$, the model is trained to associate its possible future export from every country in year $y+\delta$, to the information about the respective export baskets of all products in year $y$. The rationale is that the algorithm will base its predictions upon learning the similarity patterns between different products, using different countries as different observations. 
In the present study, we set the input data $X$ at the 2 digit aggregation level: hence, for each of the 5040 6 digit products $y$, the input is represented by the export data about the 97 2 digit aggregated productive sectors. 

\subsubsection*{Cross-validation}

Given the strong temporal auto-correlation of the network \cite{albora2021product}, the knowledge of the present export basket of a country is very informative on its future export basket. So, in order to make sure that the predictions provided by the model are based solely on its learning of the correlations between products, rather than on its ability to recognize the country, we perform a 13-fold cross-validation procedure. The 169 countries are divided into 13 groups $\bigl\{C_{k}\bigr\}_{k=1}^{13}$ of 13 countries each. For each product, we then build 13 different models, where each one is trained on data about the 156 countries $c\notin C_{k}$ and is then used to make predictions for countries $c\in C_{k}$. In this way the predictions for every country are provided by a model that did not receive any information about the country itself. 

\subsection*{Feature importance}

The directed link from a product $p$ whose presence (or absence) enhances the likelihood that a general country exports also the target product $p'$ is given by the feature importance, i.e. the relevance the RF algorithm attributes to each feature $p$ in its predictive task. The construction of each decision tree in the forest is based on the recursive split of the observations that compose the training set, in terms of the corresponding values of the features \cite{breiman2001random}: starting from the root node (containing all the observations), each node considers a feature, and depending on the binary value of this feature, the observations are divided into two child nodes. The choice of the feature for each node is meant to maximize the decrease in \emph{Gini impurity}, a metrics measuring the impurity of a node as the compresence of observations belonging to both classes (i.e. 1 and 0), given by\cite{james2013intro}:

\begin{equation*}
    G^{m}_{j} = \sum_{i=0,1}\hat{p}_{m,i}(1-\hat{p}_{m,i})
\end{equation*}
where $m$ is the node, $j$ the corresponding feature and $\hat{p}_{m,i}$ is the empirical frequency of observations in the node belonging to class $i$. The decrease in impurity realized by feature $j$ on node $m$ is then:

\begin{equation*}
    GD^{m}_{j} = G^{m} - f^{1}G^{m,1} - f^{2}G^{m,2}
\end{equation*}
where 1 and 2 indicate the two child nodes built in the split, and $f$ the corresponding fractions of observations they receive.
On a single tree $t$, being $N(j)$ the number of nodes to which feature $j$ is attached, and $V$ the total number of features, the decrease in impurity realized by feature $j$, i.e. its \emph{Gini importance}, is equal to:

\begin{equation*}
    GI_{j}(t) = \frac{\sum_{m\in N(j)}GD^{m}_{j}(t)}
    {\sum_{j=1}^{V}\sum_{m\in N(j)}GD^{m}_{j}(t)}
\end{equation*}

The \emph{Gini importance} of a feature is then given by the average decrease in \emph{Gini impurity} the feature realizes over the whole forest \cite{breiman1984cart}:

\begin{equation*}
    GI_{j} = \frac{1}{T}\sum_{t=1}^{T}GI_{j}(t)
\end{equation*}
where T is the total number of trees.

\subsubsection*{Statistical validation procedure}

Given the feature importance values, it is important to distinguish which features are actually informative for the algorithm, and which got a non-zero value because of spurious correlations in the dataset. We then implemented a statistical validation procedure similar to the one described in \cite{altmann2010permimp}, in order to compute, for each feature importance, its corresponding p-value. The method is based on the reconstruction, for each feature importance, of the corresponding \emph{null distribution}, i.e. a distribution of the importance values a feature is given by the algorithm under the hypothesis of independence between the feature itself and the response vector $y_{train}$. 

The procedure works as follows:

\begin{enumerate}
    \item for every product $p'$, we train the Random Forest 50 times and compute the \emph{Gini importance}, obtaining 50 vectors of feature importance $gi_{n}(p')$, $n=1,...,50$.
    \item we then permute the response vector $y_{train}$ 500 times, breaking its association with the feature, and recompute the \emph{Gini importance} after every permutation. In this way we obtain 500 vectors of \emph{null importance} $ni_{m}(p')$, $m=1,...,500$.
    \item for each feature, we compare each of the 50 values of \emph{Gini importance} with the 500 values of null importance: the corresponding \emph{p-value} is computed as the fraction of 500 null importance values bigger than the \emph{Gini importance} value. We then obtain, for each product, 50 vectors of \emph{p-values} $pv_{n}(p')$, $n=1,...,50$.
    \item we take the average vectors of \emph{Gini importance}:
    \begin{equation*}
        gi(p') = \frac{1}{50}\sum_{n=1}^{50}gi_{n}(p')
    \end{equation*}
    and we keep only the importance values of the features for which more than $95\%$ of the \emph{p-values} (i.e. at least 48 out of 50) are within the $95\%$ significance threshold (i.e. $p<0.05$), putting the others to 0.
\end{enumerate}

In this way we obtain, for each product, a vector containing the 97 values of statistically validated feature importance, for the 97 features. The choice of the number of repetitions and permutations is consistent with the heavy computational cost involved. It has to be noted that point 1 and point 2 are carried out separately on each of the 13 folds of the \emph{cross-validation} setting, and the corresponding values are averaged out. This computation required approximately 180 hours on a server with 20 cores

The method has been extensively tested on both low-dimensional \cite{hapfelmeier2013anew} and high-dimensional \cite{janitza2016compfast} datasets, showing a great ability to filter out the non-informative features.

\subsection*{Feature Importance Product Space}

The feature importance vectors contain information about the productive sectors recognized by the Random Forest as important in order to competitively export the corresponding product 5 years later. Therefore, the distance between the vectors relative to two products can be seen as a natural proxy for their similarity, i.e. of the overlap of capabilities needed for their export. Then, in line with the \emph{Continuous Projection Space} proposed in \cite{tacchella2021relatedness}, we project these vectors on a 2-dimensional space, via the dimensionality reduction algorithm \emph{t-SNE} \cite{van2008visualizing}: on such space, which we call \emph{Feature Importance Product Space} (FIPS), the distance between products is related to the distance between their original 97-dimensional vectors.

At this point, we can use the space to make out-of-sample forecasts on the activation of new exports after 5 years, by adopting the density-based approach explained in the following. This is a natural way to validate the FIPS idea and building procedure.\\ 
We first compute the matrix $D$ of euclidean distances between products in the FIPS. Then we transform these distances into a similarity matrix $B$, where the similarity of two products $p$ and $p'$ is computed as:

\begin{equation}
    B_{pp'} = \frac{1}{\sqrt{2\pi\sigma^{2}}}exp\Bigl[-\frac{1}{2}\Bigl(\frac{D_{pp'}}{\sigma}\Bigr)^{2}\Bigr]
\label{eq:similarity}
\end{equation}

where $\sigma$ a free parameter.\\
Given this similarity matrix, following the economic complexity literature \cite{hidalgo2007product}, we perform the prediction on $M_{cp}(y+\delta)$ by relating the likelihood of an activation to the scores defined by:

\begin{equation}
    S_{cp}(y+\delta) = \frac{\sum_{p'}B_{pp'}M_{cp'}}{\sum_{p'}B_{pp'}}
\label{eq:fips_pred}
\end{equation}

i.e., for each country $c$ the prediction on its future export of a product $p$ is given by the sum on the the products it already exports, weighted by their similarity with $p$.\\
We build the FIPS on the Random Forest trained on data in years 1996-2013, and then used it to make out-of-sample forecasting on $M_{cp}(2018)$. We use of the \emph{Python} implementation of \emph{t-SNE} algorithm provided by the library \emph{scikit-learn}\footnote{https://scikit-learn.org/stable/modules/generated/sklearn.manifold.TSNE.html}.

\subsubsection*{Optimization of the parameters}

The predictions provided by the FIPS (eq. \ref{eq:fips_pred}) depend on two parameters: the \emph{perplexity} value set for \emph{t-SNE} and the standard deviation $\sigma$ chosen for the gaussian weights (eq. \ref{eq:similarity}).
The former is a hyperparameter of the \emph{t-SNE} algorithm, fixing the expected number of elements that will be grouped into each cluster \cite{van2008visualizing}. The latter fixes the width of the gaussian distribution centered on each product to attribute the similarity weights to all the other products. The two parameters are then connected, as increasing the \emph{perplexity} value will result in a denser FIPS, and hence even for small values of $\sigma$ many products will get a high similarity score. 

\begin{figure}[ht]
\centering 
\includegraphics[width=\textwidth]{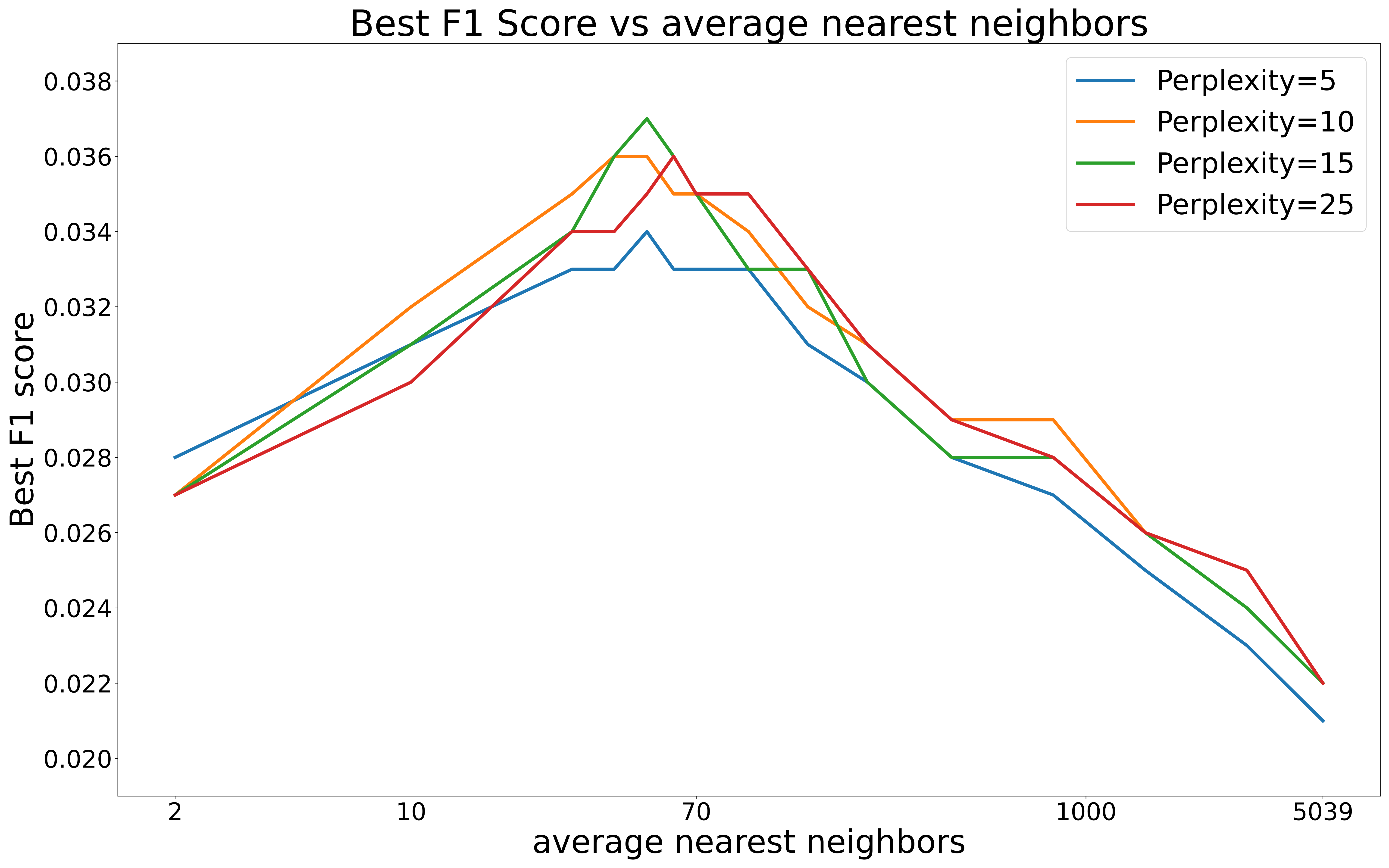}
\caption{\textbf{Best F1 Score versus average nearest neighbors for the FIPS.} We use this curve to optimize the two (related)free parameters of the FIPS. We can see how the curves for different values of \emph{perplexity} are very close and show the same trend. There is a clear peak for a number of average nearest neighbors equal to 70, which fixes also the value of $\sigma$.}
\label{fig:f1_vs_nn}
\end{figure}

\begin{figure}[ht]
\centering 
\includegraphics[width=\textwidth]{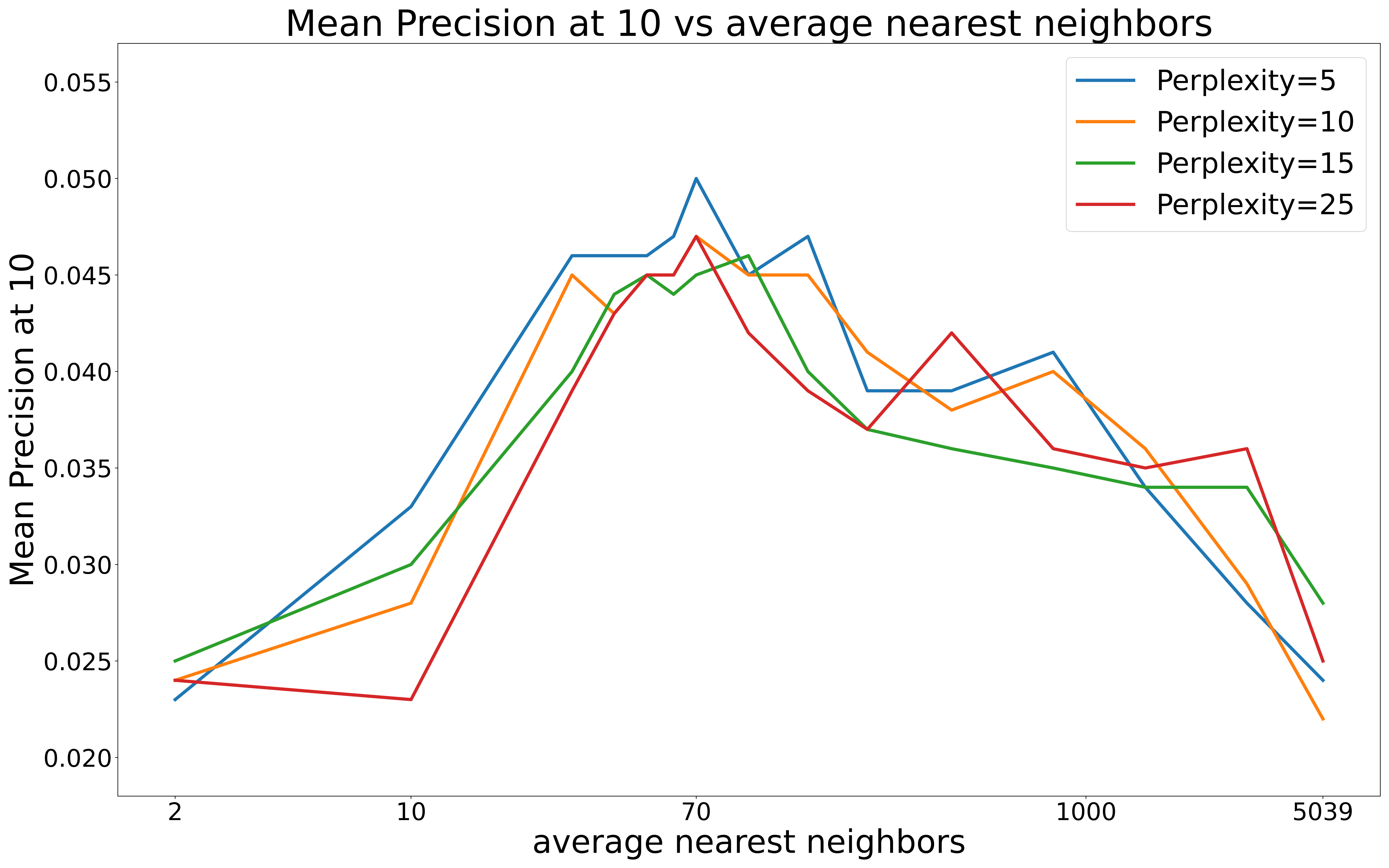}
\caption{\textbf{Mean Precision at 10 versus average nearest neighbors for the FIPS.} See previous figure. Also in this case the curves are all very close, but the trend is more noisy than for Best F1 Score. The peak is approximately at the same number of average nearest neighbors.}
\label{fig:mp@10_vs_nn}
\end{figure}

Therefore we combine them into a single parameter, which we call \emph{average nearest neighbors}, computed empirically as the (average) number of neighbouring products contained within a circle of radius $3\sigma$, centered on each product in the space. In practice, for a fixed value of \emph{perplexity}, we look for the value of $\sigma$ corresponding to integer numbers of \emph{average nearest neighbors}, and then evaluate the performance of the FIPS, measured by \emph{Best F1 Score} and \emph{Mean Precision at 10}, as a function of this number. In Fig. \ref{fig:f1_vs_nn} and \ref{fig:mp@10_vs_nn} we report the trends of the metrics for four different values of \emph{perplexity} ($perp=5,10,15,20$): for both metrics the curves are very close for different \emph{perplexity} values, showing a peak around 70 nearest neighbours. We chose to set $perplexity=10$, and the corresponding value $\sigma=4.58$. 
The performance values reported in the Results section are computed for these values of the parameters.

\clearpage

\subsection*{Logit model}

To assess the additional information carried by the FIPS with respect to the temporal auto-correlation of the $M$ matrices, we use the predictions provided by FIPS and $RCA(2013)$ as independent variables in a logistic regression for the probability of products' appearances in 2018, given by the equation:

\begin{equation}
    S^{logit}_{cp}('18) = \alpha+\beta RCA_{cp}('13)+\gamma S^{FIPS}_{cp}('18)
\end{equation}

where $S^{FIPS}_{cp}('18)$ is the FIPS prediction for the element $m_{cp}('18)$.
The prediction matrices $RCA(2013)$ and $S^{FIPS}(2018)$, as well as $M_{cp}(2018)$, are stacked into three vectors and the training set is built as:

\begin{itemize}
    \item $X_{train}=(\vec{RCA}(2013),\vec{S}^{FIPS}(2018))$
    \item $y_{train}=\vec{M}(2018)$
\end{itemize}

The model is trained only on the \emph{activations} (defined by $RCA_{cp}(y)<0.25$ for $y\in[1996-2013]$, see \cite{tacchella2021relatedness}). In order to test the out-of-sample performance, we divide the training set into 13 subsets, following a cross-validation procedure: the predictions $\vec{S}_{k}(2018)$ for each group $k$ ($k=1,...,13$) are provided by a model trained on the remaining 12, and so can be tested against the corresponding elements $\vec{M}_{k}(2018)$.

The logistic regression was carried out using the \emph{Logit} algorithm provided by the \emph{Python} library \emph{statsmodels}\footnote{https://tedboy.github.io/statsmodels\_doc/generated/generated/statsmodels.api.Logit.html}.

\subsection*{Performance metrics}

To evaluate the predictive performances of the models, we made use of a series of evaluation metrics commonly used in Machine Learning. As already mentioned, the predictions $S_{cp}(2018)$ are probability values, to be binarized in order to compare them with the answers given by the matrix elements $M_{cp}(2018)$. In order to avoid the introduction of an arbitrary binarization threshold $t$, we opted for the use of \emph{"threshold-free"} metrics, assessing the overall predictive performance of the models. Moreover, given the strong class imbalance of the dataset (the fraction of positive elements in the $M_{cp}$ matrices oscillates around the 10\% of the total elements in the covered time span, see\cite{albora2021product}), we avoided metrics such as accuracy, awarding the correct individuation of true negatives (i.e. correct classification of elements $M_{cp}(2018)=0$, which are often trivial).
The chosen metrics are:

\begin{itemize}
    \item \textbf{AUC-ROC}. The $AUC-ROC$, as suggested by the name (Area Under the Curve of the Receiving Operator Characteristic) \cite{hanley1982meaning},\cite{bradley1997use} measures the area under the Receiving Operator curve, i.e. the curve in the $TPR(t)\,vs\,FPR(t)$ plane (respectively True Positive Rate and False Positive Rate, see\cite{powers2011evaluation}) obtained by varying the value of the binarization threshold $t$. Its value, ranging from 0 to 1, represents the probability that the classifier attributes an higher score to a positive element rather than to a negative one: $AUC-ROC=1$ represents a perfect classifier, while $AUC-ROC=0.5$ corresponds to a totally random classifier. It has been shown \cite{saito2015precision} that the $AUC-ROC$ is not fully reliable when the classifier is applied to an imbalanced dataset, which is our case (see \cite{albora2021product}), as it tends to overestimate the actual accuracy of the predictions.
    \item \textbf{mean Precision at k}. The \emph{precision} is defined as the ratio between the \emph{true positives} (i.e. the positively classified elements that are actually positive) and all the positively classified elements \cite{powers2011evaluation}. We can define the \emph{Precision at k} as the \emph{precision} of the classifier on the $k$ top-ranked elements, i.e. we classify the $k$ elements with higher prediction scores as positives and then compute the corresponding \emph{precision}. The \emph{mean Precision at k} is obtained by computing the \emph{Precision at k} for every country individually, and then taking the average over all countries. Since the most diversified countries tend to activate more products than the low and medium income ones, the averaging procedure allows to filter-out this effect, retaining an overall estimate of the classifier's performance. The value of $k$ was set to 10.
    \item \textbf{Best F1 Score}. The \emph{F1 Score} is defined as the harmonic mean of \emph{precision} and \emph{recall} \cite{powers2011evaluation}. Therefore it provides an estimate of the overall quality of the classifier, as it assumes an high value only if both $precision$ and $recall$ are high. Since these two quantities rely on the choice of a binarization threshold $t$, we adopted the \emph{Best F1 Score}, i.e. the \emph{F1 Score} computed for the value of $t$ that maximizes it.
\end{itemize}

\bibliography{sciadvbib}
\bibliographystyle{ScienceAdvances}

\noindent \textbf{Acknowledgements:} 
The authors would like to thank the CREF project "Complessità in Economia".

\noindent \textbf{Data and materials availability:} The starting database about the export volumes between countries is available from UN-COMTRADE (https://comtrade.un.org), upon subscription. The processed data used in this work is available from the authors upon reasonable request.

\clearpage

\section*{Supplementary Information}

\subsection*{S1. Complexity of products on the FIPS}

As stated in the main text (see Feature Importance Product Space in the Results section), the heterogeneous structure of the FIPS seems to be related to the complexity of the different productive sectors. Specifically, sectors grouped into individual, more distinguishable clusters can be intuitively associated to a low level of industrial sophistication, while more complex sectors occupy denser and more mixed regions of the space (see Fig. 2 in the main body).
To test this qualitative intuition we coloured products in the FIPS according to their value of complexity, as computed from the Fitness algorithm \cite{tacchella2013economic}: each product was assigned the average of its (log-)complexities values in the time span 1996-2013, and the corresponding ranking was then normalized to fit into the [0,1] interval. 

\begin{figure}[ht]
\centering 
\includegraphics[width=\textwidth]{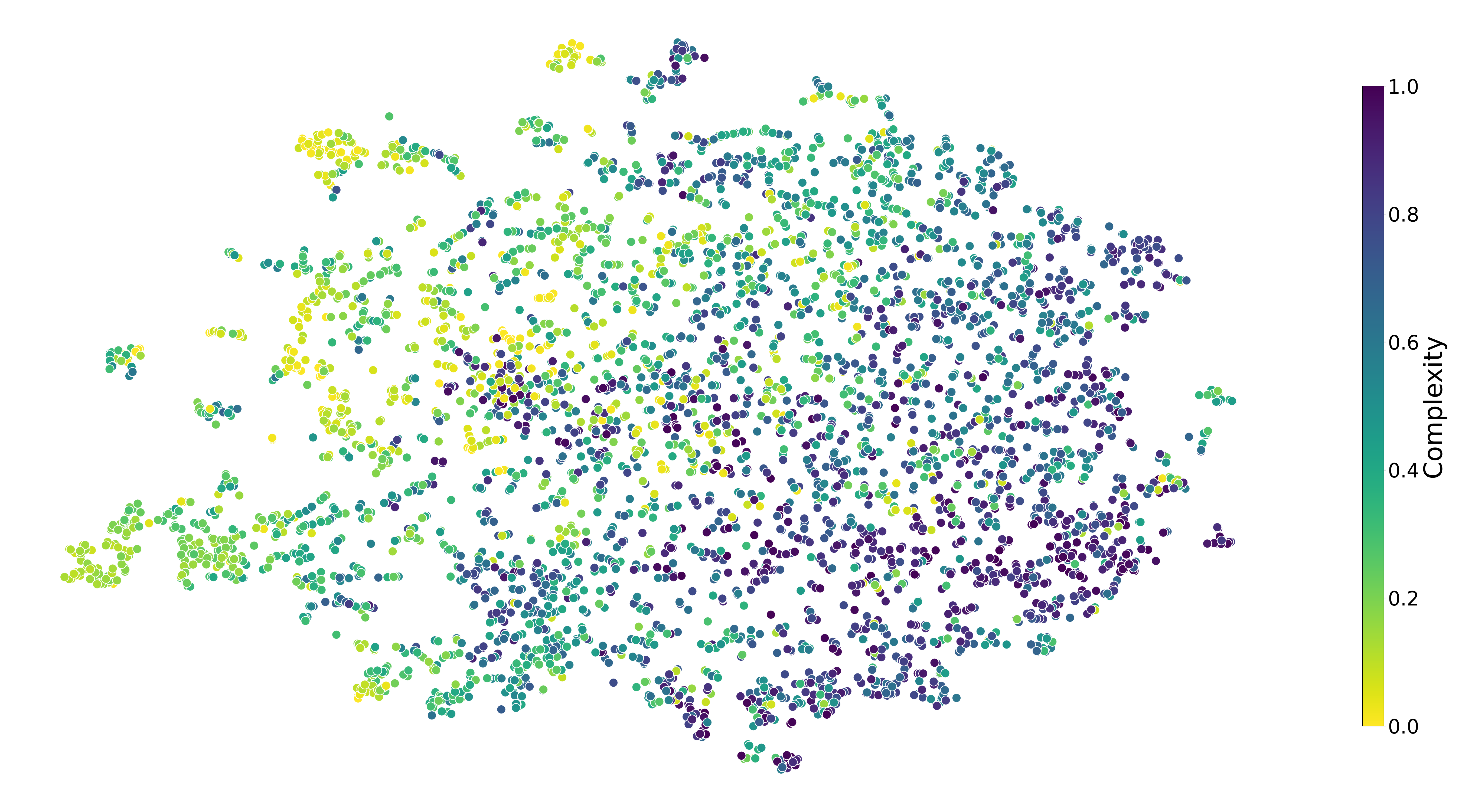}
\caption{\textbf{Feature Importance Product Space coloured according to products' complexity.} As qualitatively stated in the main text, the heterogeneous structure of the FIPS, regarding the division among productive sectors, is related to the complexity of products. The sectors making up isolated clusters (as the ones on the left side, identified as belonging Textiles and Agrifood) have low complexity products, while high complexity products live in denser regions of the space, showing a mixture of different sectors (as the right side of the FIPS). The complexity of products show an increasing pattern moving from the top-left to the bottom-right side of the space.}
\label{fig:fips_compl}
\end{figure}

The result, shown in Fig. \ref{fig:fips_compl}, confirms the expectations: clusters of products on the left side, belonging to individual productive sectors (e.g. Agrifood, Textiles and Hides and Wood, see Fig. 2 in the main body) show a low level of complexity; conversely the right region of the space, containing a mixing of different sectors (e.g. Machinery, Vehicles, Instruments, see Fig. 2 in the main body) is dominated by high-complexity products.  

\subsection*{S2. Analysis at the sector level}

In order to have an additional, qualitative test of the ability of the implemented procedure to detect significant similarity and, most importantly, necessity patterns between economic activities, we applied it to the 97 aggregated 2-digit productive sectors, rather that to the 5040 6-digit products. In details, we carried out the exact same procedure to compute the feature importance values, but aggregating the target products at the 2 digit level: i.e. for each of the 97 sectors, the Random Forest was trained to associate its export in $y+\delta$ to the export of all the sectors in year $y$. 

\subsubsection*{Feature Importance Product Network}

In this "symmetric" setting, by stacking vertically the feature importance vectors, we built a $97\times97$ feature importance matrix $F$, where each element $F_{ss'}$ represents the importance sector $s$ has for the future export of sector $s'$ (the target). Putting the diagonal elements to 0, we discarded trivial terms corresponding to the sectorial self-contributions. 

This matrix can be interpreted as the adjacency matrix of a weighted directed network, with links going from each sector to the ones it is important for, and weights given by the values of this importance. Hence we decided to inspect this network, to determine whether its topological structure reflects intuitive patterns of similarity and necessity relationships among sectors. 

In order to visualize the main structural features of the network in a clear fashion, we applied the filtering method known as $PMFG$ (Planar Maximally Filtered Graph), introduced by Tumminello et al. in \cite{tumminello2005tool}. The method consists in the iterative addition to the empty graph consisting of just the nodes, of the links with the highest weights, retaining only the ones that keep the graph "planar"\cite{west2001introduction}: the resulting network consists of $3(N-2)$ edges, and contains the Minimum Spanning Tree\cite{gower1969minimum} of the original network, representing its first extension \cite{tumminello2005tool}).
Since the $PMFG$ does not account for multi-edges and directions, as a preliminary step we discarded multi-edges between sectors retaining, for every couple of sectors, the link with the highest weight. Directions were added back manually after the filtering procedure.

\begin{figure}[hb]
\centering 
\includegraphics[width=\textwidth]{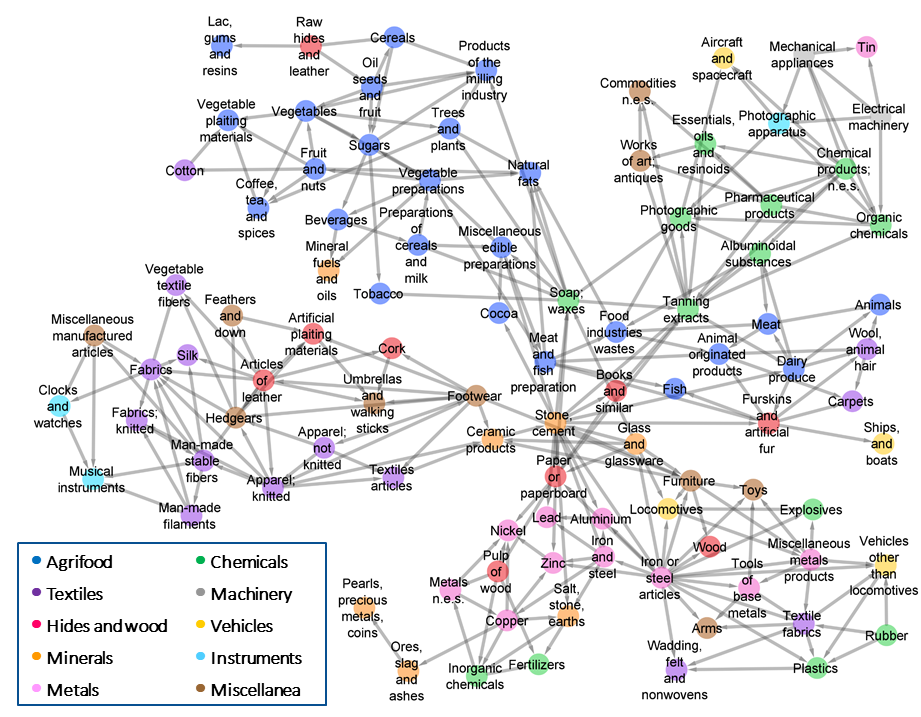}
\caption{\textbf{Feature Importance Product Network.} Visualization of the (filtered) network constructed from the feature importance matrix: the links signal feature importance relations among sectors. Products are coloured according to their reference macro-category (see Tab. \ref{tab:macrocategories}). The graph displays a clear division in clusters that mostly reflect intuitive associations between different productive sectors.}
\label{fig:fipn}
\end{figure}

The result, which we call FIPN (Feature Importance Product Network), is reported in Fig. \ref{fig:fipn}: sectors are coloured according to their macro-category (see Tab. \ref{tab:macrocategories}). The topological structure of the network confirms the ability of the method to learn significant dependence patterns between sectors, as the clusters reflect intuitive associations among different sectors, even belonging to different macro-categories. At the top-left of the network there is a cluster of agricultural products, while animal-related products form a different, but close, cluster on the right. On the left side there is a cluster of products belonging to the apparel industry, while in the top-right region there is a group of chemical products, together with 
sectors related to heavy industry. Finally, at the bottom of the network there is a huge cluster grouping different kind of building materials (metals, minerals, wood), chemicals, vehicles and mixed manufactured products.

It is to be noted that the FIPN shows also several counter-intuitive edges, e.g. the one linking "furskins and artificial fur" to "ships and boats": this odd results, along with the limitations of the method, have to be attributed to the high level of noise and spurious correlations encoded in the data at this high level of aggregation. 

\clearpage

\subsubsection*{Feature importance and products' complexity}

\begin{figure}[hb]
\centering 
\includegraphics[width=\textwidth]{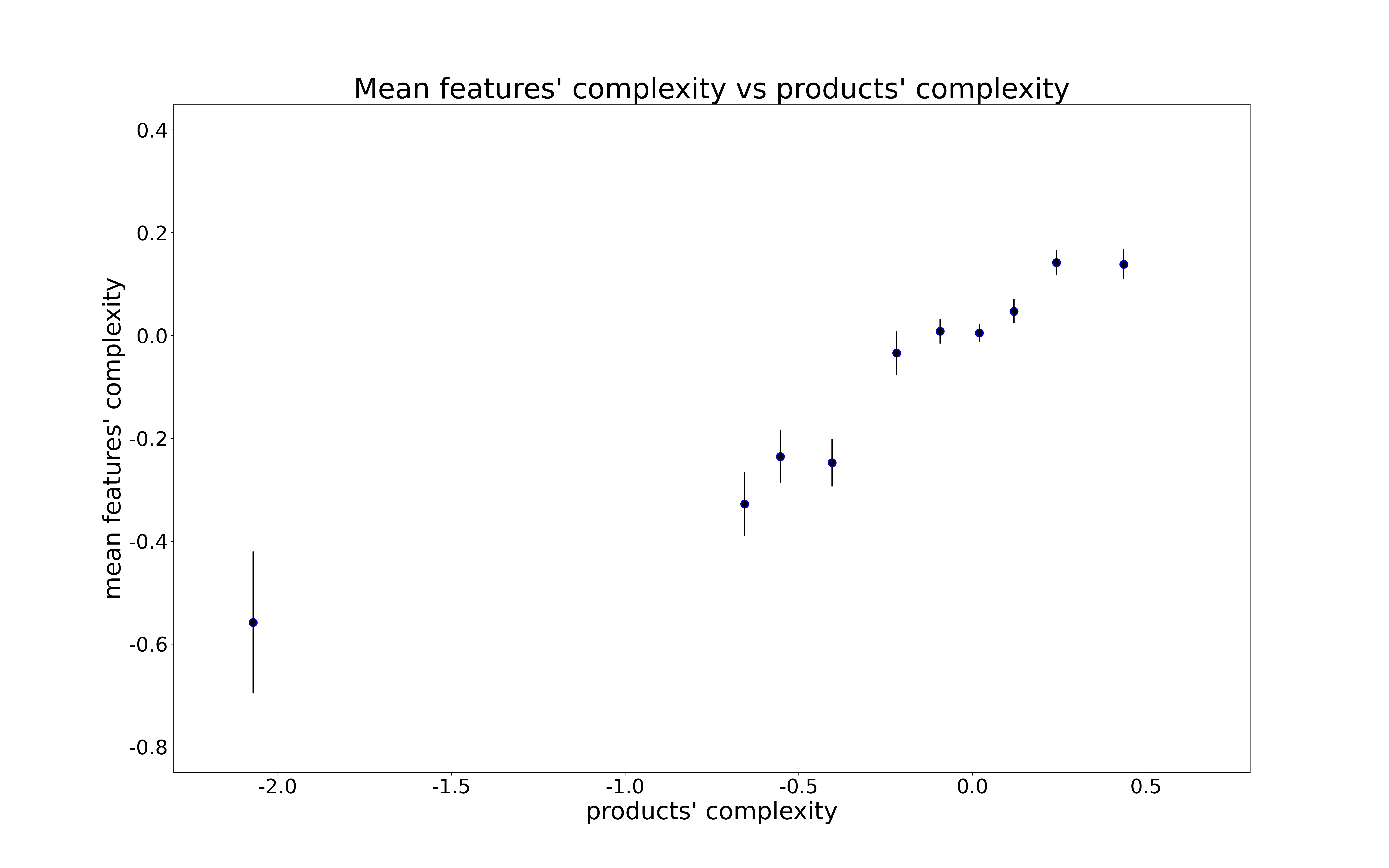}
\caption{\textbf{Mean complexity of features versus mean complexity of sectors.} In this case the trend is in line with the expectations, as the mean complexity of features increases with increasing complexity of sectors in the whole interval. The reported values of complexities are computed as the average of the logarithms of complexities in years 1996-2013. The sectors are grouped into 10 bins of 9-10 sectors each, for which we show the average mean complexity of features and the corresponding standard error.}
\label{fig:fimp_compl_vs_compl_2dig}
\end{figure}

Regarding the connection between feature importance and the complexity of sectors, we carried out the same analysis done for the 6-digit products (see Results), plotting the mean complexity of validated features for each sector versus its complexity. The complexity of each sector is computed as the average of the (log-)complexities in the 1996-2013 interval. The curve, reported in Fig. \ref{fig:fimp_compl_vs_compl_2dig}, shows a trend perfectly in line with the one observed in the 6 digit case (see Fig. 3 in the main text). The average complexity of features grows for increasing complexity of the corresponding product, in the whole interval, confirming that more complex products need, on average, more complex features in order to be competitively exported.

\clearpage

\subsection*{S3. Table of products' macro-categories}

In Table \ref{tab:macrocategories} we report the composition of the productive macro-categories employed in the paper.

\begin{table}[!hb] 
\centering 
\medskip 
\begin{tabular}{ll}
\toprule
\multicolumn{2}{c}{\textbf{Agrifood}}\\
\hline
\midrule
\textbf{01}: Live animals  & \textbf{13}: Lac, gums and resins \\
\textbf{02}: Meat  & \textbf{14}: Vegetable plaiting materials \\
\textbf{03}: Fish & \textbf{15}: Natural fats 
\\
\textbf{04}: Dairy produce & \textbf{16}: Meat and fish preparations 
\\
\textbf{05}: Animal originated products  & \textbf{17}: Sugars \\
\textbf{06}: Trees and plants  & \textbf{18}: Cocoa 
\\
\textbf{07}: Vegetables  & \textbf{19}: Preparations of cereals and milk 
\\
\textbf{08}: Fruit and nuts  & \textbf{20}: Preparations of vegetables 
\\
\textbf{09}: Coffee, tea and spices  & \textbf{21}: Miscellaneous edible preparations 
\\
\textbf{10}: Cereals  & \textbf{22}: Beverages 
\\
\textbf{11}: Products of the milling industry  & \textbf{23}: Food industries wastes 
\\
\textbf{12}: Oil seeds and fruit  & \textbf{24}: Tobacco
\\
\hline
\multicolumn{2}{c}{\textbf{Minerals}}\\
\hline
\midrule
\textbf{25}: Salt, earths  & \textbf{68}: Stone, cement \\
\textbf{26}: Ores, slag and ash  & \textbf{69}: Ceramic products \\
\textbf{27}: Minerals fuels and oils & \textbf{70}: Glass and glassware 
\\
\textbf{71}: Pearls, precious metals, coins &
\\
\hline
\multicolumn{2}{c}{\textbf{Chemicals}}\\
\hline
\midrule
\textbf{28}: Inorganic chemicals  & \textbf{29}: Organic chemicals
\\
\textbf{30}: Pharmaceutical products  & \textbf{31}: Fertilisers \\
\textbf{32}: Tanning or dyeing extracts & \textbf{33}: Essential oils and resinoids
\\
\textbf{34}: Soap, waxes & \textbf{35}: Albuminoidal substances
\\
\textbf{36}: Explosives & \textbf{37}: Photographic goods
\\
\textbf{38}: Chemical products n.e.s. & \textbf{39}: Plastics
\\
\textbf{40}: Rubber & 
\\
\hline
\multicolumn{2}{c}{\textbf{Hides and wood}}\\
\hline
\midrule
\textbf{41}: Raw hides and leather  & \textbf{42}: Articles of leather 
\\
\textbf{43}: Furskins and artificial fur  & \textbf{44}: Wood 
\\
\textbf{45}: Cork & \textbf{46}: Artificial plaiting materials 
\\
\textbf{47}: Pulp of wood & \textbf{48}: Paper and paperboard
\\
\textbf{49}: Books and similar &  
\\
\bottomrule
\end{tabular}
\end{table}

\begin{table}[!hb] 
\centering 
\medskip 
\begin{tabular}{ll}
\toprule
\multicolumn{2}{c}{\textbf{Textiles}}\\
\hline
\midrule
\textbf{50}: Silk  & \textbf{51}: Wool, animal hair 
\\
\textbf{52}: Cotton  & \textbf{53}: Vegetable textile fibers 
\\
\textbf{54}: Man-made filaments & \textbf{55}: Man-made staple fibers
\\
\textbf{56}: Wadding, felt and nonwovens & \textbf{57}: Carpets
\\
\textbf{58}: Fabrics & \textbf{59}: Textile fabrics
\\
\textbf{60}: Fabrics, knitted & \textbf{61}: Apparel, knitted.
\\
\textbf{62}: Apparel, not knitted & \textbf{63}: Textile articles
\\
\hline
\multicolumn{2}{c}{\textbf{Metals}}\\
\hline
\midrule
\textbf{72}: Iron and steel  & \textbf{73}: Iron or steel articles 
\\
\textbf{74}: Copper  & \textbf{75}: Nickel 
\\
\textbf{76}: Aluminium  & \textbf{78}: Lead
\\
\textbf{79}: Zinc  & \textbf{80}: Tin 
\\
\textbf{81}: Metals n.e.s.  & \textbf{83}: Tools of base metals 
\\
\textbf{83}: Miscellaneous metals products  &
\\
\hline
\multicolumn{2}{c}{\textbf{Machinery}}\\
\hline
\midrule
\textbf{84}: Mechanical appliances  & \textbf{85}: Electrical machinery
\\
\hline
\multicolumn{2}{c}{\textbf{Vehicles}}\\
\hline
\midrule
\textbf{86}: Locomotives  & \textbf{87}: Vehicles other than locomotives
\\
\textbf{88}: Aircraft and spacecraft  & \textbf{89}: Ships and boats
\\
\hline
\multicolumn{2}{c}{\textbf{Instruments}}\\
\hline
\midrule
\textbf{90}: Photographic apparatus  & \textbf{91}: Clocks and watches
\\
\textbf{92}: Musical instruments  & 
\\
\hline
\multicolumn{2}{c}{\textbf{Miscellanea}}\\
\hline
\midrule
\textbf{64}: Footwear  & \textbf{65}: Hedgears
\\
\textbf{66}: Umbrellas and walking sticks  & \textbf{67}: Feathers and down
\\
\textbf{93}: Arms  & \textbf{94}: Furniture
\\
\textbf{95}: Toys  & \textbf{96}: Miscellaneous manufactured articles
\\
\textbf{97}: Works of art  & \textbf{99}: Commodities n.e.s.
\\
\bottomrule
\end{tabular}
\caption{\textbf{Productive macro-categories.} Composition of the productive macro-categories in terms of the 2 digit productive sectors.}
\label{tab:macrocategories}
\end{table}

\clearpage

\end{document}